\documentclass[modern]{aastex62}

\newcommand{\nicer}{\textit{NICER}{}}

\graphicspath{{./}{figures/}}

\shorttitle{Unusual burst oscillations in 4U~1728}
\shortauthors{Mahmoodifar et al.}

\begin{document}

\title{\textit{NICER} Observation of Unusual Burst Oscillations in 4U~1728$-$34}

\correspondingauthor{Simin Mahmoodifar}
\email{siminmahmoodifar@gmail.com}

\author[0000-0003-2386-1359]{Simin Mahmoodifar}
\affil{Astrophysics Science Division, NASA Goddard Space Flight Center, Greenbelt, MD 20771, USA}
\affil{Department of Astronomy, University of Maryland, College Park, MD 20742, USA}
\author[0000-0001-7681-5845]{Tod E. Strohmayer}
\affil{Astrophysics Science Division and Joint Space-Science Institute, NASA Goddard Space Flight Center, Greenbelt, MD 20771, USA}
\author{Peter Bult}
\affil{Astrophysics Science Division, NASA Goddard Space Flight Center, Greenbelt, MD 20771, USA}
\author[0000-0002-3422-0074]{Diego Altamirano}
\affil{Physics \& Astronomy, University of Southampton, Southampton, Hampshire SO17 1BJ, UK}
\author{Zaven Arzoumanian}
\affil{Astrophysics Science Division, NASA Goddard Space Flight Center, Greenbelt, MD 20771, USA}
\author{Deepto Chakrabarty}
\affil{MIT Kavli Institute for Astrophysics and Space Research, Massachusetts Institute of Technology, Cambridge, MA 02139, USA}
\author{Keith C.~Gendreau}
\affil{Astrophysics Science Division, NASA Goddard Space Flight Center, Greenbelt, MD 20771, USA}
\author[0000-0002-6449-106X]{Sebastien Guillot}
\affil{IRAP, CNRS, 9 avenue du Colonel Roche, BP 44346, F-31028 Toulouse Cedex 4, France}
\affil{Universit\'{e} de Toulouse, CNES, UPS-OMP, F-31028 Toulouse, France}
\author[0000-0001-8371-2713]{Jeroen Homan}
\affil{Eureka Scientific, Inc., 2452 Delmer Street, Oakland, CA 94602, USA}
\affil{SRON, Netherlands Institute for Space Research,
    Sorbonnelaan 2, 3584 CA Utrecht, The Netherlands}
\author[0000-0002-6789-2723]{Gaurava K. Jaisawal}
\affil{National Space Institute, Technical University of Denmark, Elektrovej 327-328, DK-2800 Lyngby, Denmark}
\author{Laurens Keek}
\affil{Department of Astronomy, University of Maryland, College Park, MD 20742, USA}
\author[0000-0002-4013-5650]{Michael T. Wolff}
\affil{Space Science Division, U.S. Naval Research Laboratory, Washington, DC 20375, USA}



\begin{abstract}

The Neutron Star Interior Composition Explorer (\nicer) has observed seven thermonuclear X-ray bursts from the Low Mass X-ray Binary (LMXB) neutron star 4U~1728$-$34 from the start of the mission's operations until February of 2019. Three of these bursts show oscillations in their decaying tail with frequencies that are within 1 Hz of the previously detected burst oscillations from this source. Two of these burst oscillations have unusual properties: They have large fractional rms amplitudes of $ 48 \pm 9 \%$ and $ 46 \pm 9 \%$, and they are detected only at photon energies above 6 keV. By contrast, the third detected burst oscillation is compatible with previous observations of this source, with a fractional rms amplitude of $7.7 \pm 1.5\%$ rms in the 0.3 to 6.2 keV energy band. We discuss the implications of these large-amplitude burst oscillations, finding they are difficult to explain with the current theoretical models for X-ray burst tail oscillations. 

\end{abstract}



\section{Introduction} \label{sec:intro}
Type I X-ray bursts are thermonuclear explosions on the surfaces of Low Mass X-ray Binary (LMXB) neutron stars that are due to unstable burning of material, predominantly hydrogen and helium, accreted onto their surfaces \citep[for recent reviews]{Watts2012,Galloway2017}. X-ray bursts are detected as short-lived flashes of emission from a neutron star, where the X-ray flux rises by a factor of up to 15 in a few seconds and decays back to the persistent flux level over tens to hundreds of seconds.
Nearly coherent, periodic variations in the X-ray flux of a bursting neutron star, so-called burst oscillations, have been observed in less than 20\% of all accreting neutron stars that exhibit type I X-ray bursts. Even for those 20\% of sources, only some bursts from each object show oscillations. Burst oscillations in each source always occur at frequencies that are within a few hertz of each other. Also, when oscillations are observed during the rise of a burst, the fractional amplitude of oscillations is always larger at the start of the burst and decreases to a minimum near the peak. Using these observations, \cite{Strohmayer1996} and others concluded that burst oscillations are rotationally induced modulations that are caused by an asymmetric temperature distribution on the neutron star surface. This interpretation was confirmed by the discovery of burst oscillations from the accreting millisecond X-ray pulsar SAX J1808.4--3658 \citep{Chakrabarty2003}, which demonstrated that burst oscillation frequencies closely match the spin frequency of the neutron star \citep{Wijnands1998}. The detection of burst oscillations may therefore be used to measure the neutron star spin frequency in non-pulsating sources.
The frequency of burst oscillations evolves during the rise and decaying tail of a burst and it usually asymptotes to the spin frequency of the pulsar during the tail.
The fractional root mean square (rms) amplitudes of burst oscillation signals are usually between 5--25\% \citep{Galloway2008}. There are some cases where larger amplitudes have been observed in the rising phase of a burst. For example, \cite{Strohmayer1998} reported a fractional half-amplitude of $75\% \pm 17\%$ (an rms amplitude of $\sim 53\%$) in a short interval at the beginning of a burst from 4U~1636$-$54; however, burst oscillations in the tail generally have smaller amplitudes, with an average of about 10\% \citep{Galloway2008}.  

Burst oscillations in the rise can be explained by the spreading of a burning region on the surface of a rotating neutron star; however, understanding the cause of oscillations in the tail has been more challenging. Several models have been proposed to explain the oscillations in the decaying tail of X-ray bursts. \cite{Heyl2004} suggested that they can be due to oscillation modes with low frequencies on the surface of a neutron star, such as surface r-modes. However, theoretical estimates for the frequencies of such modes in the corotating frame of the neutron star overestimate the separation between burst oscillation frequencies and stellar spin frequencies measured in pulsars \citep{Berkhout2008, Muno2002}. Alternatively, \cite{Cumming2000} suggested that the tail oscillations might be due to a cooling wake, the temperature asymmetry due to the cooling of the neutron star surface. This is because it takes a finite time for both the atmosphere to cool and the burning to spread around the star. \cite{Spitkovsky2002} showed that the temperature gradient will drive a zonal thermal wind moving opposite to the star's rotation, and suggested that if an inhomogeneous feature such as a vortex were trapped in it, this could produce a flux modulation. \cite{Mahmoodifar2016} proposed a phenomenological model with asymmetric cooling, where the speed of the cooling wake is different in different regions on the star, and is not symmetric about the rotation axis. In this model the temperature contrast between the hot and cold regions on the stellar surface can produce high amplitude oscillations in the tail, consistent with the highest observed, but lacks a first-principle explanation of how such an asymmetry might arise. Still more work is needed in order to understand the cause of large-amplitude oscillations observed in the tails of some bursting objects.

4U~1728$-$34 (hereafter, 4U~1728) is a well-known, X-ray bursting LMXB \citep{Lewin1976, Basinska1984} that has been classified as an atoll source based on observations with the EXOSAT satellite \citep{Hasinger1989}. Burst oscillations were first detected in this source shortly after the launch of the Rossi X-ray Timing Explorer ({\it RXTE}) \citep{Strohmayer1996}. Many subsequent X-ray bursts detected with {\it RXTE} showed oscillations with frequencies between 361 and 364 Hz in their rise and/or tail. The frequency of these oscillations exhibited an upward trend, usually starting at a lower value in the rise and asymptoting to a higher value in the tail. The fractional rms amplitudes of these oscillations were always less than 25\% in the rise and less than 15\% in the tail \citep{Franco2001,vanStraaten2001,Galloway2008}.

In this paper we discuss the \textit{Neutron Star Interior Composition Explorer} (\nicer; \citealt{Gendreau2016}) observations of several type I X-ray bursts from 4U~1728, and report on the detection of burst oscillations in three of them. We focus primarily on two bursts that show unusual oscillations with very high amplitudes during their decaying tails.

\section{Observations}
\label{sec:obs}
\nicer{} is a soft X-ray telescope that is mounted on the International Space Station. It consists of 56 co-aligned X-ray concentrator optics, each paired with a silicon-drift detector; 52 of these detectors are operating, providing an effective area of $\sim1900$ cm$^2$ at 1.5 keV. \nicer{} is sensitive in the 0.2--12 keV energy band, with energy resolution better than 150 eV and time resolution of $\sim 100$ ns rms.

\nicer{} has observed seven type I X-ray bursts from 4U~1728 since the start of its operations (see Table \ref{table:bursts}). We have detected burst oscillations in three of these bursts. The first occurrence of burst oscillations was on 2017 September 2 (hereafter, burst 4), the second one on 2018 February 7 (hereafter, burst 6), and the third one on 2018 July 9 (hereafter, burst 7). Here we focus mostly on the first and third detections, as they showed highly unusual properties (see below). We have analyzed the data using version 1.02 of the \nicer{} response files.

\begin{deluxetable}{ccccccc}
\tablecaption{\nicer{} observations of 4U~1728$-$34 X-ray bursts}
\label{table:bursts}
\tabletypesize{\scriptsize}
\tablecolumns{7}
\tablehead{
\colhead{Number} & \colhead{ObsID} & \colhead{Start Time} & \colhead{Date} & \colhead{Peak Count-Rate} & \colhead{Oscillations} & \colhead{Chance Probability} \\
\colhead{} & \colhead{} & \colhead{(MJD)} & \colhead{(YYYY-MM-DD)} & \colhead{(ct/s)} & \colhead{} & \colhead{(Single Trial)}
} 
\startdata
1 & 0050150106 & 57940.82556 & 2017-07-06 & 2400 & & \tabularnewline
2 & 1050150102 & 57953.11803 & 2017-07-19 & 3000 & & \tabularnewline
3 & 1050150111 & 57979.45593 & 2017-08-14 & 3000 & &\tabularnewline
4 & 1050150127 & 57998.10078 & 2017-09-02 & 2500 & Yes & $6.5\times 10^{-9}$\tabularnewline
5 & 1050150134 & 58006.66959 & 2017-09-10 & 2250 & & \tabularnewline
6 & 1050150149 & 58156.69746 & 2018-02-07 & 2500 & Yes & $2.8\times 10^{-7}$\tabularnewline
7 & 1050150158 & 58308.89837 & 2018-07-09 & 2300 & Yes & $1.02\times 10^{-7}$\\
\enddata
\tablecomments{\nicer{} observation ID, date of observation, peak count rate, whether or not burst oscillations are observed in each burst from 4U~1728$-$34 and the single trial chance probability of candidate signals.}
\end{deluxetable}

\begin{figure}
\centering
\includegraphics[width=5 in]{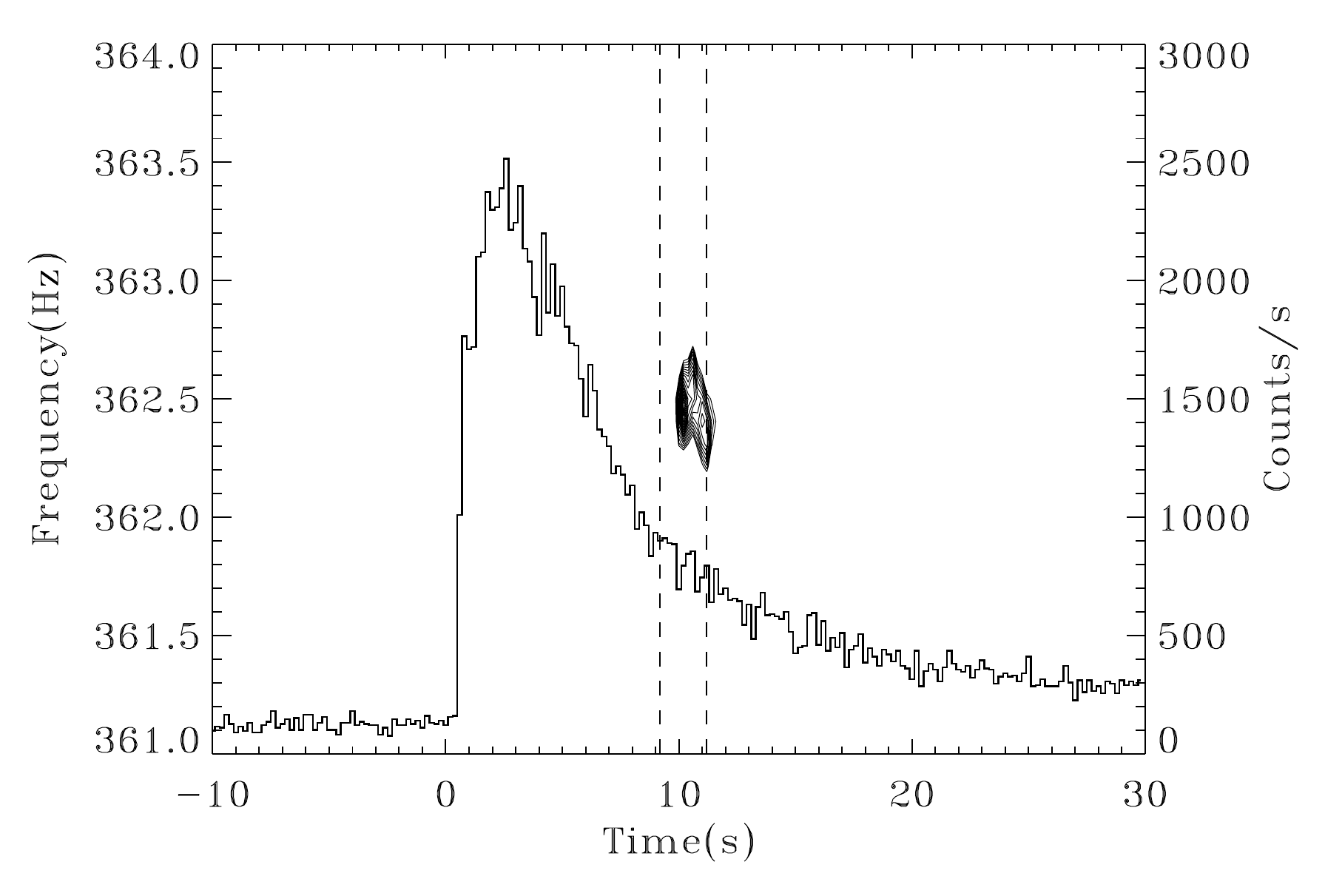}
\caption{The dynamical power spectrum overplotted on the \nicer{} light curve of burst 4 from 4U~1728. The light curve (right-hand vertical axis) is computed using $0.3 \leq E \leq 12$ keV photons and bin size of 0.2 s. The dynamical power spectrum (left-hand vertical axis) is computed from overlapping 2 s intervals, with a new interval starting every 0.2 s, using 6--12 keV photons. Contours are plotted for Leahy-normalized power values of 20 to 37, in steps of 1. \label{fig:Lc}}
\end{figure}

\begin{figure}
\centering
\includegraphics[width=2.8in]{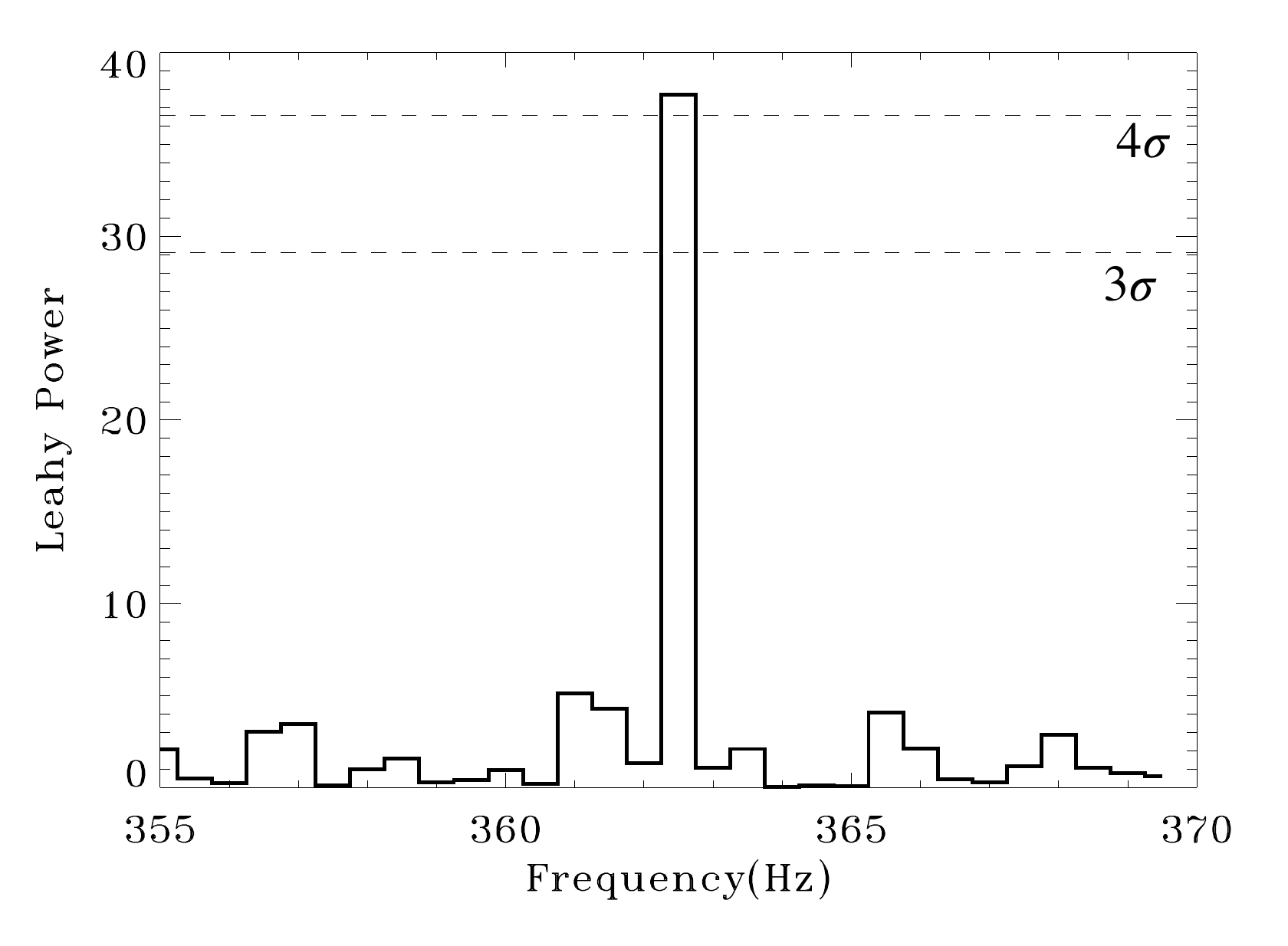} 
\includegraphics[width=2.95in]{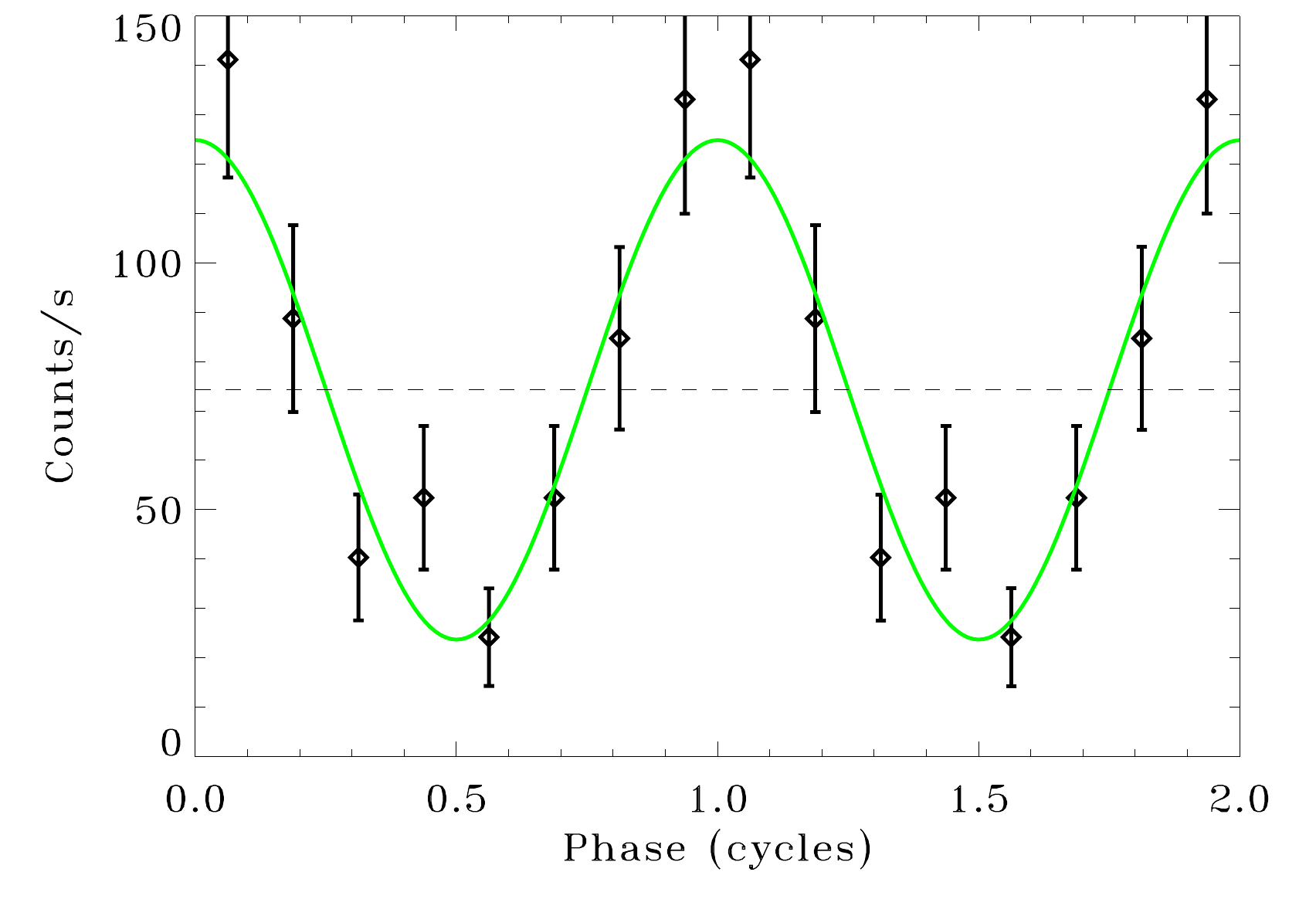}
\caption{Left panel: Leahy normalized power spectrum computed from the 2 s interval during the tail of burst 4, shown in Fig.~\ref{fig:Lc}. The power spectrum is computed using the events in the 6.2--9.9 keV energy range, which maximizes the power. The two horizontal dashed lines show the 3 and 4 sigma levels, counting all the trials in our search. Right panel: Pulse profile in the 6.2--9.9 keV band obtained by folding the 2 s interval indicate in Fig.~\ref{fig:Lc} at the oscillation frequency of 362.5 Hz in 8 phase bins. Two cycles are plotted for clarity. The best fitting sinusoidal model (solid curve), $A + B \sin(\phi - \phi_0)$, is also plotted. The horizontal dashed line shows the value $A=74.3$. \label{fig:power}}
\end{figure}

\subsection{Burst Oscillation Analysis}
\label{sec:timing}
\subsubsection{Timing Analysis}

In this section we discuss our method of searching for burst oscillations during each burst in Table \ref{table:bursts}. We constructed Leahy-normalized \citep{Leahy1983} dynamical power spectra for the time interval starting approximately 5 s before the burst onset and spanning 37 s. The dynamical power spectrum consists of individual power spectra, each covering a 2 s interval. Each new interval overlaps with the previous one and starts 0.2 s  later than the previous one. We restricted our search window to frequencies between 360--365 Hz, which covers the observed range for burst oscillations in 4U~1728 \citep{vanStraaten2001}. Initially, we searched three different energy bands, 0.3--12 keV, 0.3--6 keV, and 6--12 keV. Each dynamical power spectrum contains 175 individual power spectra, which in turn have $2\times 5=10$ frequency bins; because we searched in three energy ranges, the total number of trials in our dynamical search is 5250. 
We then focused on the time interval where the highest peak was detected and attempted to maximize the power by varying the search parameters. We examined 10 energy cuts ($10 \times 10=100$ extra trials) and also a few time intervals, each 3 or 4 s long, that included the original 2 s interval. This added an extra 370 trials to the original 5250 trials from the dynamical power spectrum, which made the total number of trials in each burst 5620\footnote{Because power spectra are computed in overlapping time intervals, their frequency bins are not fully independent; nevertheless, for the purpose of estimating the significance of each signal we count them as independent trials. In order to understand the effect of these correlations we have done  Monte Carlo simulations of our search procedure that are discussed later in this section.}. 

We carried out this procedure on all of the bursts shown in Table 1 and found candidate burst oscillation signals in three of them. In burst 4 we found a peak at $f=362.5$ Hz with more than 3 sigma significance in the high energy band (6--12 keV) and during the cooling tail of the burst. The power was maximized in the energy range of 6.2 to 9.9 keV. We did not detect any significant signal in the soft band.
Fig.~\ref{fig:Lc} shows the dynamical power spectra overplotted on the X-ray burst lightcurve of this burst. The lightcurve represents the 0.3--12 keV band and the dynamical power spectrum shown here was computed in the 6--12 keV band. Our search interval started at $t=-5$ s on this plot and extended to $t=32$ s.  Fig.~\ref{fig:power}, left panel, shows the power spectrum computed for the 2 s interval shown by the two vertical dashed lines in Fig.~\ref{fig:Lc}. Note that we only searched in the frequency range of 360 to 365 Hz, around the expected frequency for burst oscillations in this source, but in this figure we show a wider range of frequencies for illustrative purposes. The power spectrum is computed in the 6.2 to 9.9 keV energy band. The maximum power is at a frequency of 362.5 Hz with a Leahy normalized value of $\sim 37.7$, which has a single-trial chance probability of $6.5 \times 10^{-9}$, and  taking into account all trials the signal has a significance of $\sim 4.1$ sigma. The two horizontal dashed lines show the 3 and 4 sigma levels counting all the trials.

Burst 6 was the second burst in which we detected burst oscillations. The left panel of Fig.~\ref{fig:Lc_amplitude49} shows the dynamical power spectrum overplotted on the light curve of this burst. The highest signal power in our search procedure was found to be in a 3 s interval during the tail of the burst, indicated by two vertical dashed lines, and it was maximized in the 0.3 to 6.2 keV energy band. The maximum power occured at a frequency of $\sim 362.77$ Hz with a Leahy normalized value of $\sim 30.2$, which has a single trial chance probability of $2.8 \times 10^{-7}$ and considering all 5620 trials it is $\sim 3.16$ sigma significant. The dynamical power spectrum shown in this plot is computed using 3 s long windows. Note that for the purpose of this plot we computed the dynamical power spectrum with 3 s windows, but in our search procedure we used 2 s windows. An increasing trend in frequency during the 3 s interval, by approximately 0.4 Hz, is evident. 

The third instance of burst oscillations was detected in burst 7, in which we found oscillations at 363.67 Hz during the tail. The power was maximized in a 4 s interval and in the 6--12 keV energy band, with a Leahy normalized value of 32.19. The single trial noise probability for this detection is $1.02\times 10^{-7}$ and taking into account all trials in our search its significance is 3.44 sigma. The lightcurve and dynamical power spectrum for this burst are shown on the left panel of Fig.~\ref{fig:Lc_amplitude58}. The light curve is computed in the 0.3--12 keV band and the dynamical power spectrum is computed in the 6--12 keV band. 

\subsubsection{Simulations of Search Procedure}

We note that the trials in each of these bursts are not independent, as they were constructed from overlapping data segments (overlapping time intervals in making the power spectra as well as overlapping energy bands). In order to more accurately quantify our signal detection significance in the presence of such correlated trials, we carried out Monte Carlo simulations that closely follow our search procedure for detecting a signal. We simulated lightcurves with the same mean count rate as our observed lightcurves in the three energy bands (0.3--12, 0.3--6 and 6--12 keV). We sample the burst flux variability using 1/8 s bins. We generated Poisson realizations by measuring the mean rate every 1/8 sec and using the same time bin size and lightcurve duration as observed.  We then computed dynamical power spectra as described earlier and computed power spectra at the same 10 frequencies between 360 and 365 Hz. Our simulations closely follow the actual search procedure. We repeated this process many times to determine how often a single trial probability as small as the ones observed in the real data is achieved in at least one of the lightcurves. We found that the significance values from the simulations are smaller than our estimated values, by a factor of $\sim 1.5$, which is not unexpected given the fact that the trials are not all independent.

We also note that although we are searching multiple bursts (seven) for oscillations, we want to test individual bursts and not the burst sample jointly.  Because it is known that individual bursts in this source can show oscillations and we also know the frequency range of the oscillations, it is appropriate to search each burst individually. Indeed, if one used all the trials from the sample (7$\times$5620) for searching each burst, then it seems plausible that one could miss a real signal in an individual burst. However, one can also carry out a joint search, effectively asking the question of whether or not the full sample of seven bursts is consistent with the null hypothesis of assuming no signals are present in any bursts.   We used our simulations to assess this question as well. We ran Poisson simulations as discussed above for the seven bursts taken together. In this case, a single ``trial" of the simulation is a Poisson realization for all seven bursts.  We then searched for any trials where at least three of the seven bursts in a single iteration showed 3 sigma detections or better. Because this requires generating many more simulations we restricted the total number of iterations to 10,000.   We found that none of these 10,000 simulated trials yielded at least three such detections (at 3 sigma or better).  This further supports our conclusion that at least some of the bursts do have real oscillation signals. And while each individual burst may have a modest detection significance, the probability of getting three out of seven such detections by chance is exceedingly low.

\subsubsection{Amplitude Measurements}

In order to compute the fractional amplitude of oscillations in each burst we phase-fold the corresponding light curve in the time interval of interest, and then fit the phase-folded light curve with the model $A + B \sin(2\pi\nu t - \phi_0)$, and express the oscillation amplitudes in terms of fractional rms, which is defined as $\vert B \vert /(\sqrt{2}A)$.


Note that in noisy data one will always get a positive value for the pulsed amplitude even in the absence of a real signal. In order to estimate the 99\% upper limits on the amplitudes in the low energy band we need to take into account the effect of noise. Following \citet{Groth1975}, \citet{Vaughan1994}, and \citet[see their equations 6 to 10]{Watts2005}, if the true signal power is $P_{sig}$, then the measured values $P_m$ will be distributed according to
\begin{equation}
p_n(P_m : P_{sig}) =\frac{1}{2} \text{exp}\left[-\frac{(P_m + P_{sig})}{2}\right]\left(\frac{P_m}{P_{sig}}\right)^{(n-1)/2} \times I_{n-1}\left(\sqrt{P_mP_{sig}}\right),
\end{equation}

\noindent where the function $I_{n-1}$ is a modified Bessel function of the first kind and order $n-1$. Here $p_n(P_m : P_{sig})$ is the true probability distribution of measured (total) power $P_m$ for the sum of $n$ independent power spectral bins containing noise and signal power $P_s$. The probability of obtaining a measured power that lies between 0 and $P_m$, given $P_{sig}$, is obtained through the associated cumulative distribution function:

\begin{equation}
f_n(P_m : P_{sig}) =\int_0^{P_{m}}p_n(x : P_{sig})dx= 1 - \text{exp}\left[-\frac{(P_m + P_{sig})}{2}\right] \times \left[\Sigma_{k=0}^{\infty} \Sigma_{l=0}^{k+n-1} \frac{(P_{sig})^k (P_m)^l}{l! k! 2^{k+l}}\right]
\end{equation}

The probability of the true signal power lying between 0 and $P_s$ given a measured power $P_m$ is then given by

\begin{equation}
f_n(P_s : P_m)=1 - f_n(P_m : P_s)
\end{equation}

Using these equations we can find the signal power $P_s$ for which $f_n(P_s : P_m)=0.99$ using different measured powers and determine the upper limits on the amplitude using $A_{rms}\approx \sqrt{P_s/N}$, where $N$ is the total number of counts in the profile, and $n=1$ in our case.

\begin{deluxetable}{cccc}
\tablecaption{Amplitudes of 4U~1728$-$34 Burst oscillations}
\label{table:bursts}
\tabletypesize{\scriptsize}
\tablecolumns{4}
\tablehead{
\colhead{Number} & \colhead{ObsID} & \colhead{Fractional rms Amplitude} & \colhead{Energy Band} 
} 
\startdata
4 & 1050150127 & $48 \pm 9 \%$ & 6.2--9.9 keV \tabularnewline
6 & 1050150149 & $7.7 \pm 1.5 \%$ & 0.3--6.2 keV \tabularnewline
7 & 1050150158 & $46 \pm 9 \%$ & 6--12 keV \\
\enddata
\tablecomments{\nicer{} observation ID for 4U~1728$-$34 bursts with observed oscillations, fractional amplitude of burst oscillations, and the corresponding energy band where the signal was maximized.}
\end{deluxetable}

The right panel of Fig.~\ref{fig:power} shows the pulse profile of burst 4 in the 6.2--9.9 keV band obtained by folding the 2 s interval of the light curve, specified by dashed lines in Fig.~\ref{fig:Lc}, at the oscillation frequency of 362.5 Hz in 8 phase bins. Two cycles are plotted for clarity. Note that there are only 153 counts in the 6.2--9.9 keV band in that 2 s interval. The best fitting model of $A + B \sin(2\pi\nu t - \phi_0)$ with $A=74.3$, $B=50.6$, and $\phi_0=0.501\pi$ is shown by the green curve. The fit is good, with a minimum chi-squared of 5.66 for 5 degrees of freedom (8 phase bins $-$ 3 model parameters). The rms amplitude of the burst oscillation signal in this burst is $48 \pm 9 \%$, which is much larger than previous measurements of burst oscillation amplitudes during the tail of an X-ray burst using {\it RXTE} observations. The largest fractional rms amplitudes for oscillations detected in 4U~1728 burst tails are less than 15\% \citep{vanStraaten2001,Galloway2008}. The 99\% upper limit on the rms amplitude of burst oscillations in the 0.3--6.2 keV energy band is $\sim 7\%$, in the 0.3--3 keV band is $\sim 14\%$ and in the 3--6.2 keV band is $\sim 10\%$. Here we should note that the amplitudes reported in previous studies have been computed in different energy bands, and not necessarily for the interval that maximizes the signal power. For example van Straaten et al. (2001) have used the whole energy band of the {\it RXTE}/PCA in their burst oscillations analysis.

\begin{figure}[btp]
\centering
\includegraphics[width=3 in]{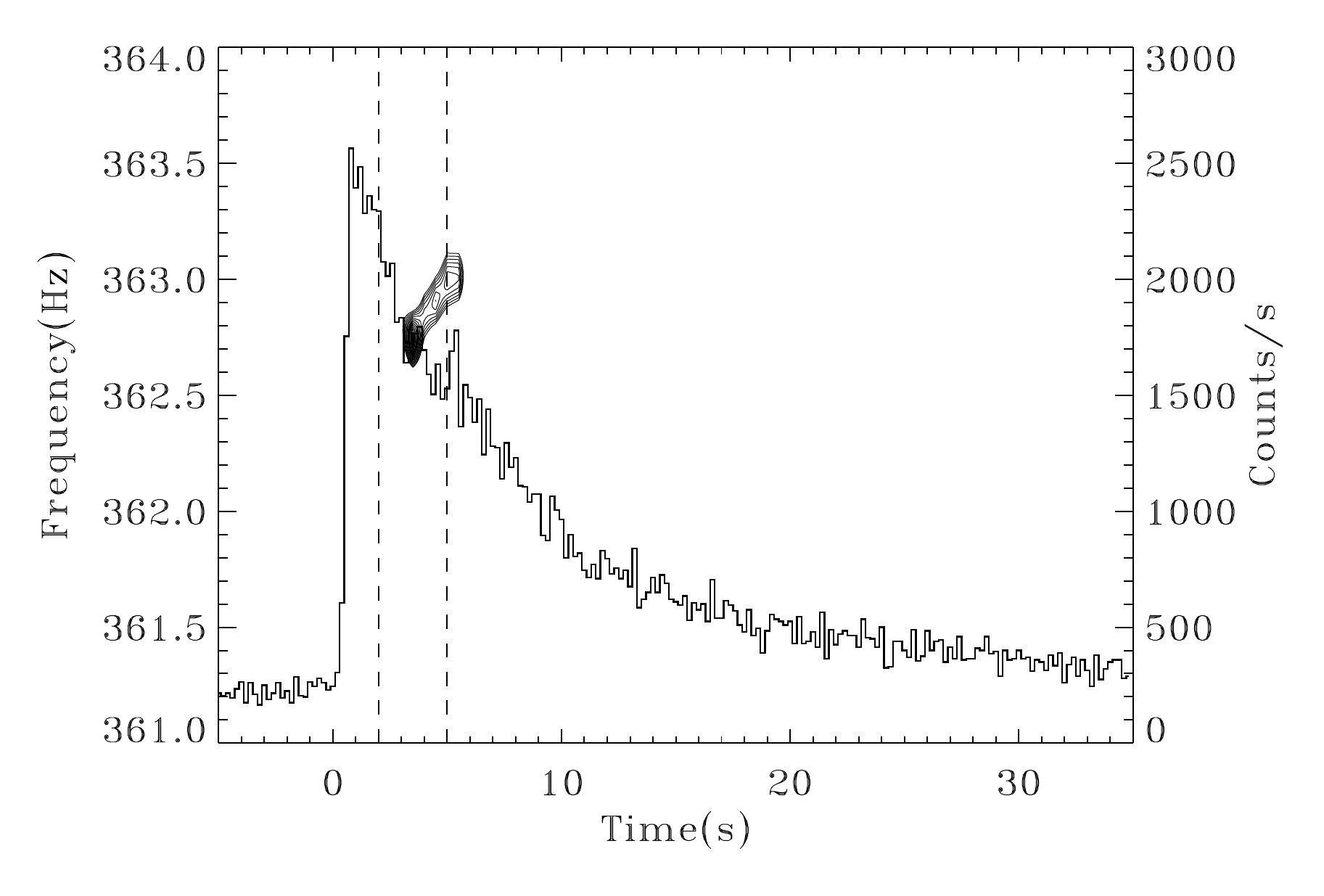} 
\includegraphics[width=2.9in]{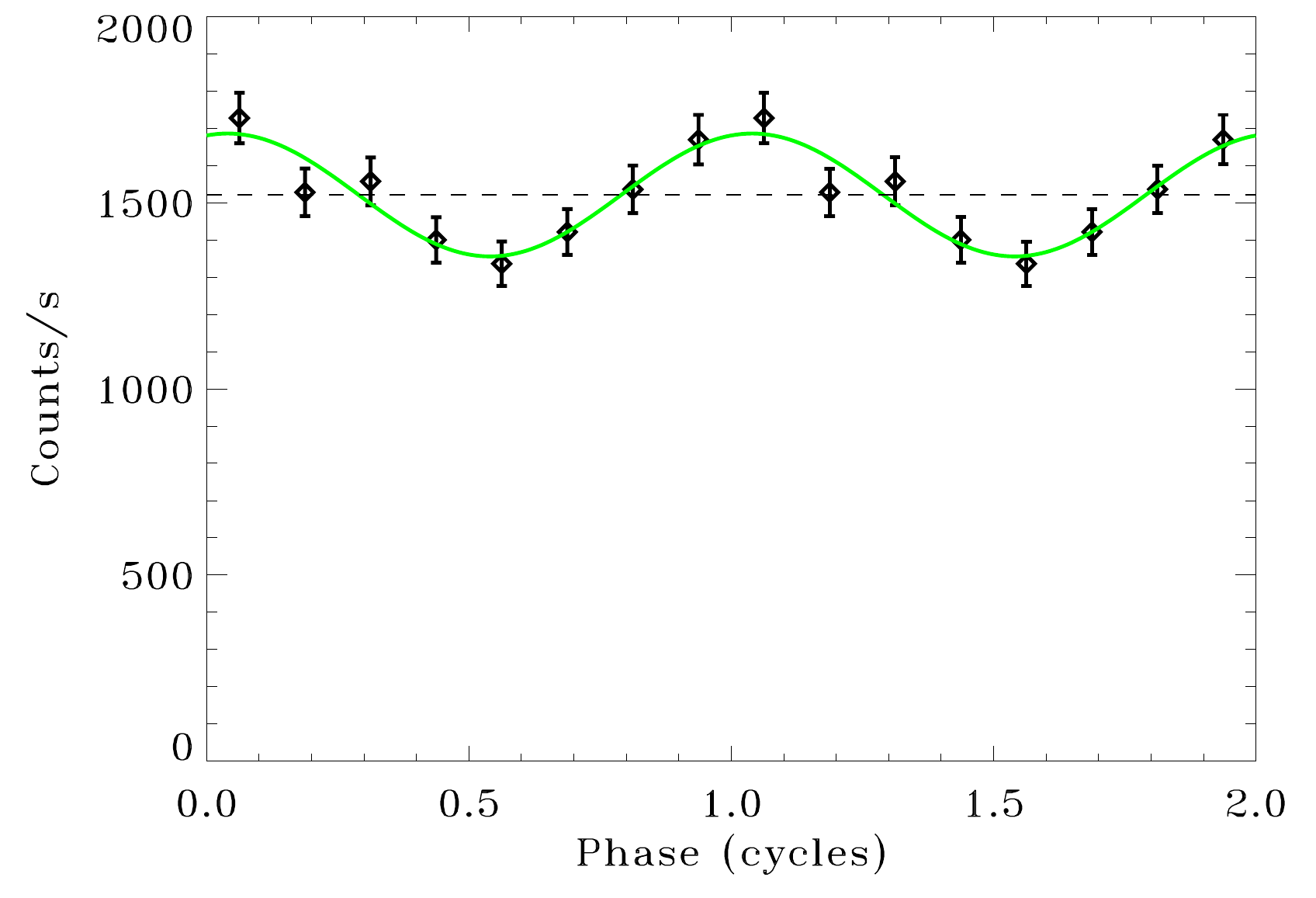} \\
\caption{Left panel: The dynamical power spectrum overplotted on the \nicer{} light curve of burst 6 from 4U~1728. The light curve is computed in the 0.3--12 keV band with a bin size of 0.2 s. The power spectra are computed using 3 s long windows, and in the 0.3--6 keV energy band. Contours are plotted for power values of 17 to 30, in steps of 1. Right panel: Pulse profile in the 0.3--6.2 keV band obtained by folding the 3 s interval specified on the light curve at the oscillation frequency in 8 phase bins. Two cycles are plotted for clarity. The best fitting sinusoidal model (solid curve), $A + B \sin(\phi - \phi_0)$, is also plotted.}
\label{fig:Lc_amplitude49}
\end{figure}

The right panel of Fig.~\ref{fig:Lc_amplitude49} shows the folded light curve for burst 6 computed from the 3 s interval shown on the left panel in the 0.3--6.2 keV energy band. The light curve is folded at the peak frequency of 362.77 Hz in 8 phase bins. The fractional rms amplitude of the burst oscillation in this case is $7.7 \pm 1.5 \%$, which is consistent with the earlier measurements of tail amplitudes using {\it RXTE} observations of 4U~1728 bursts. The parameter values of our best fitting model are $A=1521.23$, $B=165.12$, and $\phi_0=-0.42\pi$, and this fit has a minimum chi-squared value of 3.58 for 5 degrees of freedom.

\begin{figure}[btp]
\centering
\includegraphics[width=3 in]{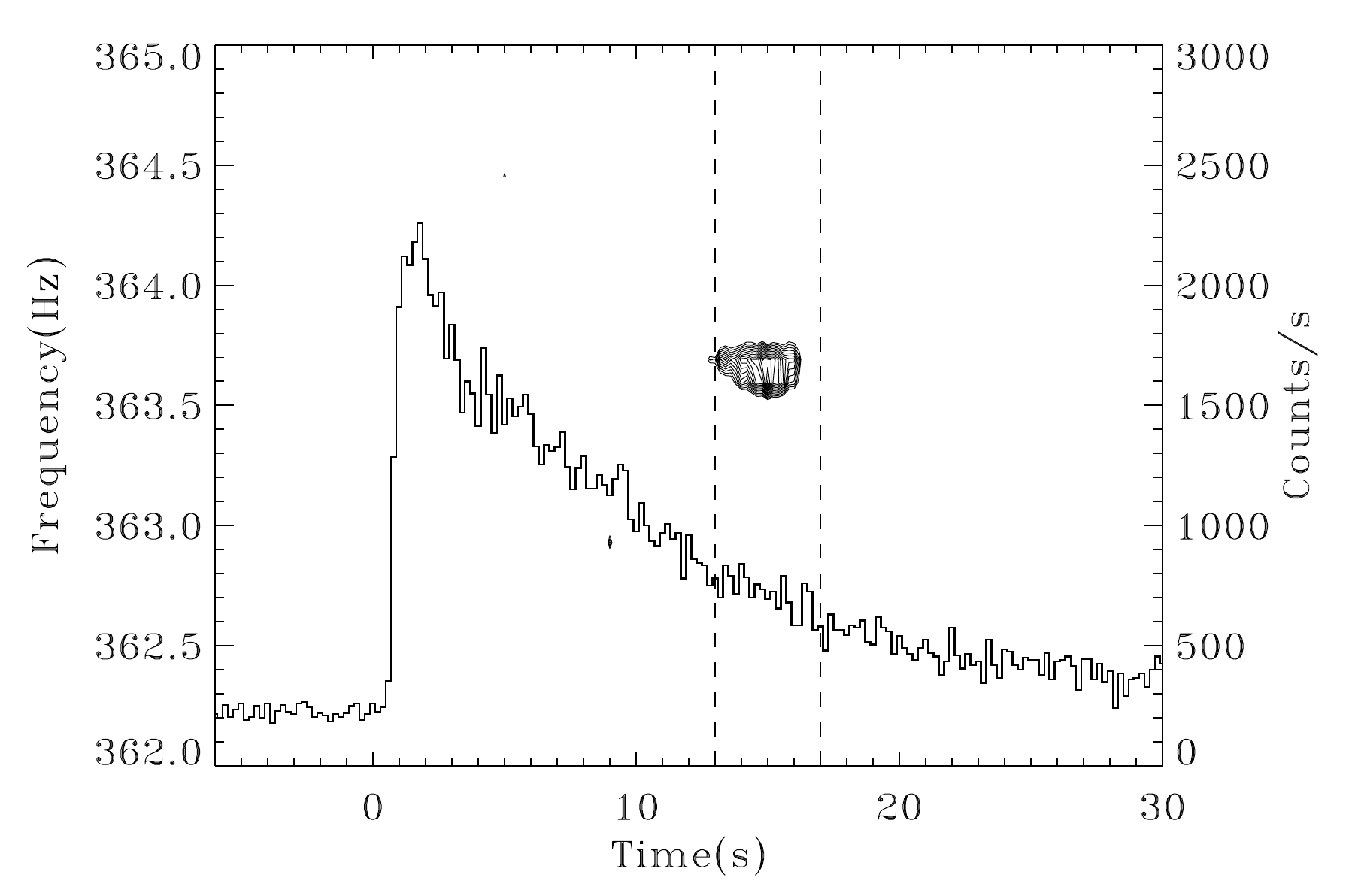} 
\includegraphics[width=2.95in]{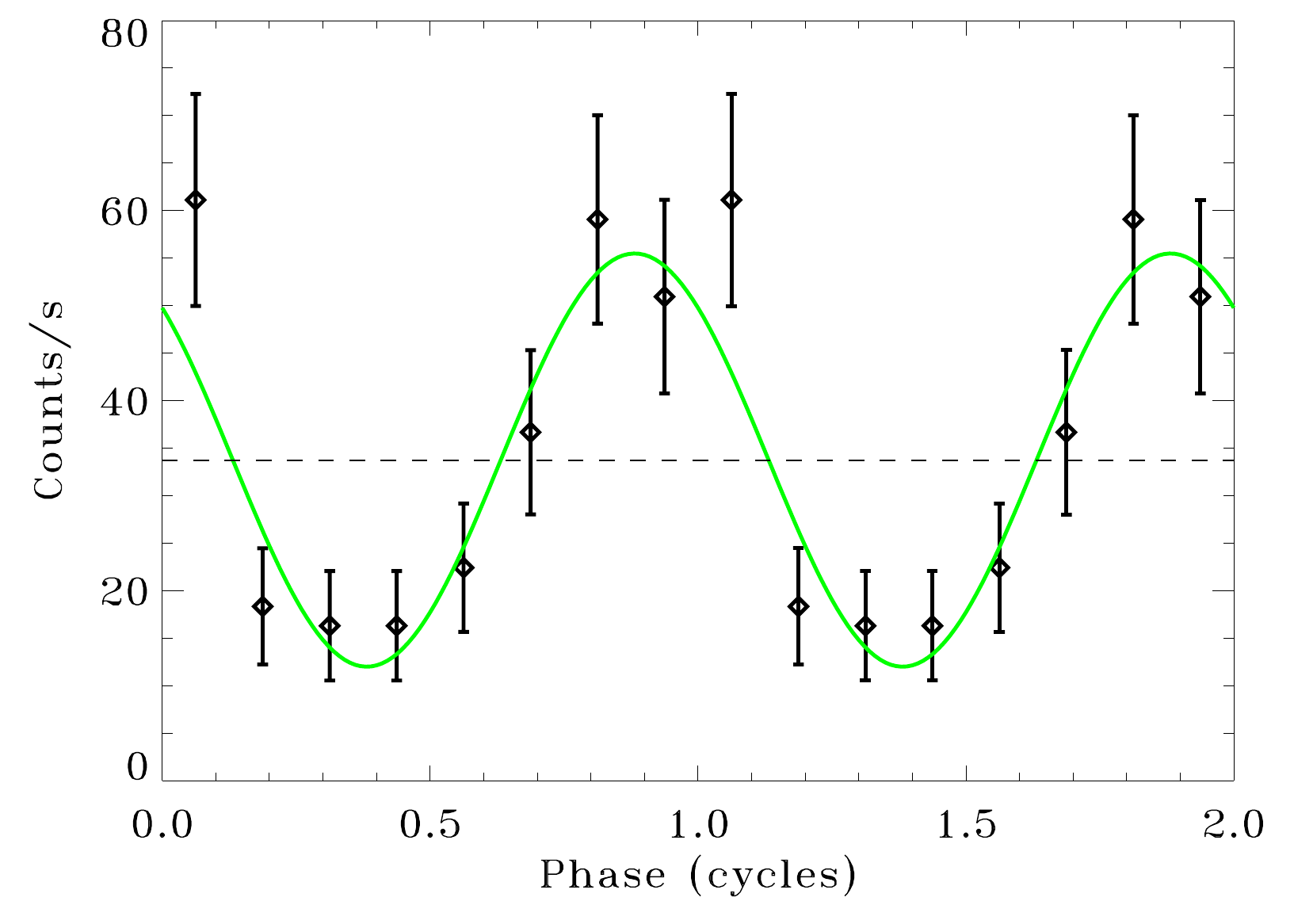} \\
\caption{Left panel: The dynamical power spectrum overplotted on the \nicer{} light curve of burst 7 from 4U~1728. The light curve is computed in the 0.3--12 keV band with a bin size of 0.2 s. The power spectrum is computed using 4 s long windows, and in the 6--12 keV energy band. Contours are plotted for power values of 18 to 32, in steps of 1. Right panel: Pulse profile in the 6--12 keV band obtained by folding the 4 s interval specified on the light curve at the oscillation frequency in 8 phase bins. Two cycles are plotted for clarity. The best fitting sinusoidal model (solid curve), $A + B \sin(\phi - \phi_0)$, is also plotted. The horizontal dashed line shows the value $A=33.7$.}
\label{fig:Lc_amplitude58}
\end{figure}

For burst 7 two cycles of the folded light curve and the best fitting model for the 4 s interval with burst oscillations are shown on the right panel of Fig.~\ref{fig:Lc_amplitude58}. The best fit parameter values in this case are $A=33.74$, $B=-21.75$, and $\phi_0=0.264\pi$, and the fit has a minimum chi-squared value of 5.52 for 5 degrees of freedom. As with burst 4, we did not detect oscillations in the 0.3--6 keV band. The fractional rms amplitude of oscillations in the 6--12 keV band is $46 \pm 9 \%$.


\section{Discussion}

Since the start of its operation, \nicer{} has observed seven type I X-ray bursts from 4U~1728 in $\sim 100$ ks of exposure. We have analyzed these data searching for burst oscillations, and detected oscillations in three of them,
all during the decaying tail, with frequencies very close to each other and consistent with those seen in previous observations of burst oscillations from 4U~1728.

\begin{figure}
\centering
\includegraphics[width=6 in]{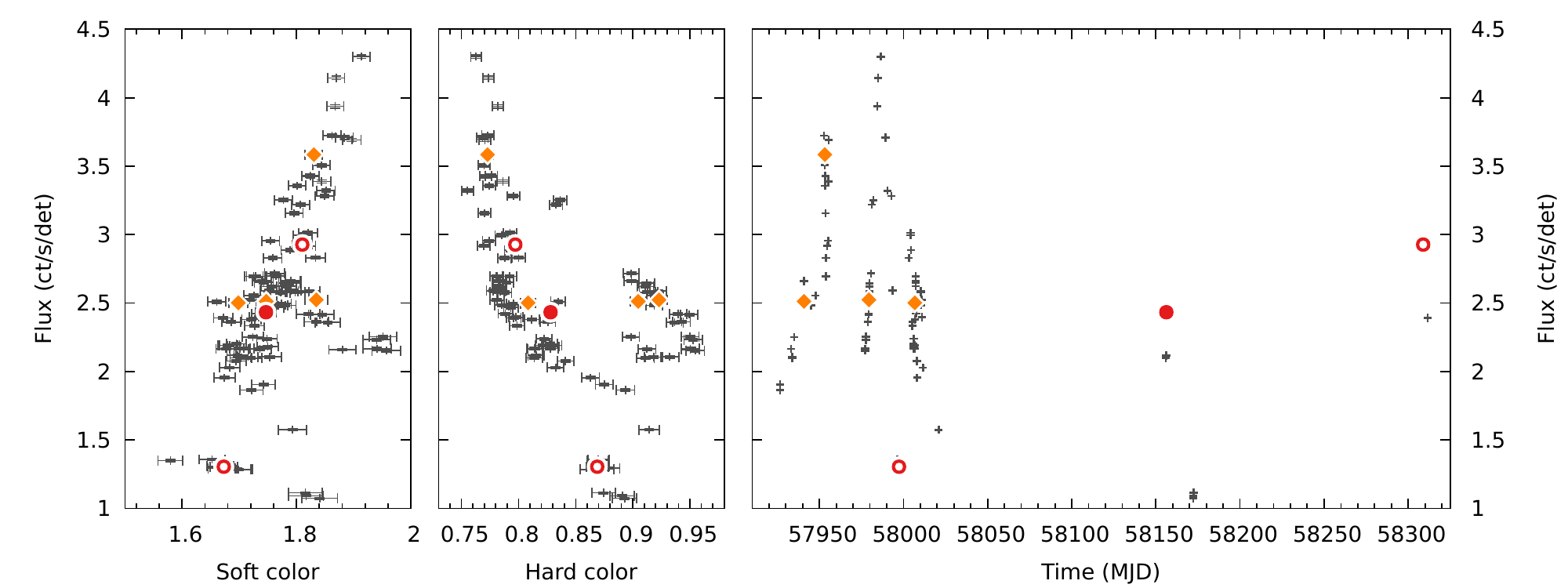}
\caption{ Panels from left to right show intensity vs.\ soft-color, intensity vs.\ hard-color, and the light curve for all \nicer{} data from 4U~1728 up to ObsID 1050150161. Each point represents a 500-s bin. All X-ray bursts are marked on these plots with red circles or orange diamonds. The ones with burst oscillations are shown in red circles. The two open circles show the bursts with high fractional amplitudes. Soft color is defined as the counts ratio in (1.8--2.5 keV)/(0.5--1.8 keV), and hard color in (3.8--6.8 keV)/(2.5--3.8 keV). \label{fig:Light_curve}}
\end{figure}

There are two distinct features of the oscillations in bursts 4 and 7: they are both detected only in the high energy band ($E > 6$ keV), and they have very large fractional rms amplitudes of $48 \pm 9 \%$ and $46 \pm 9 \%$, respectively, which has never been seen during the decaying tail of any bursts before. In order to achieve such large amplitudes one would need a large temperature contrast on the surface of the star that is confined in a small region on the star and near the equator. For example, it seems unlikely that small-amplitude oscillation modes, such as r-modes, on the surface of a neutron star could produce such large-amplitude modulations in X-ray flux. According to Narayan \& Cooper (2007), the observed flux variations in burst oscillations due to nonradial modes is expected to be about 10\%. Also, since the modes identified by \cite{Heyl2004} occupy only a small region near the equator, one would need huge pulsation amplitudes for these modes in order to produce large amplitude X-ray flux modulations. It is not clear how these modes can have such large amplitudes, and what would excite them to large amplitude during the burst tail. 
Canonical cooling wake models in which all regions on the neutron star surface heat and cool in the same manner also cannot produce large enough temperature asymmetries to explain such large amplitudes \citep{Mahmoodifar2016}.

Another physical process that can substantially affect pulsation amplitudes is beaming. Whereas existing models for burst oscillations assume isotropic surface emission, or perhaps with modest limb darkening associated with atmospheric physics, emission that is more strongly beamed toward the surface normal could produce an increased pulsed amplitude \citep{Bogdanov2014}. Another phenomenon that could perhaps be related to these observations is that of intermittent pulsations \citep{Altamirano2008,Casella2008}---the accretion-powered pulsations in some accreting millisecond pulsars (AMXPs) are known to exhibit amplitude variations that can render the pulsations undetectable at times.  Moreover, there have been some indications that the pulsed amplitude in intermittent pulsars could be influenced by the occurrence of X-ray bursts \citep{Galloway2007,Patruno2008}.  A curious feature of the two bursts with very large tail amplitudes reported here is that the oscillations are evident at quite late times in the decaying tail. Compare, for example, the relative locations of pulsations in the burst tails in Figures 3 and 4.  Indeed, the contribution of the burst to the total flux is no longer dominant at such late times in the band (6--10 keV) in which the large amplitude oscillations are detected, suggesting that perhaps these pulsations are related to the persistent emission, but are somehow enabled or induced by the burst emission.  Based on this we speculate that these oscillations may be related to a localized variation in the accretion process induced by the bursts, perhaps triggering localized infall of plasma onto the stellar surface. Strongly anisotropic beaming might result if, for example, infalling plasma was thermalized fairly high in the surface layers \citep{Mushtukov2018,Titarchuk2002,Altamirano2008}. 

In order to further study the energy dependence of the amplitude we divided the 6.2--9.9 keV range in burst 4 into two parts with almost equal counts in each one. The resulting energy bands are 6.2--7 keV (with 76 counts) and 7--9.9 keV (with 77 counts). Then we folded the light curves at the burst oscillation frequency and fitted the pulse profiles with the same model as before. The fractional rms amplitudes in the 6.2--7 keV and 7--9.9 keV bands are $\sim 54\% \pm 13\%$ and $\sim 42\% \pm 13\%$, respectively. Therefore, considering the error bars, we don't find any significant energy dependence for the amplitudes between these two energy intervals. 

\begin{figure}
\centering
\includegraphics[width=5 in]{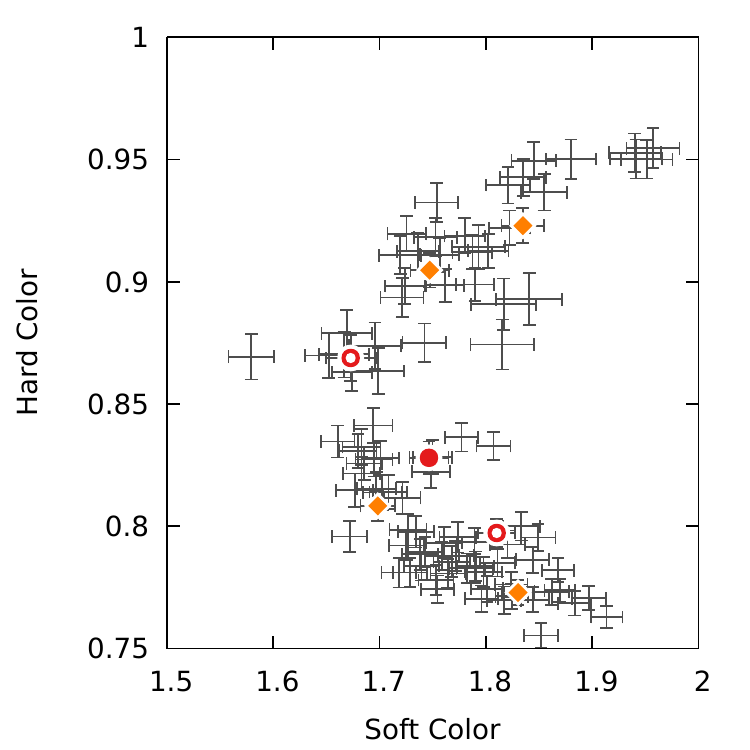}
\caption{ Color-color diagram for all \nicer{} data from 4U~1728 up to ObsID 1050150161. All X-ray bursts are marked with red or orange symbols. The ones with burst oscillations are shown in red. The two open circles show the bursts with high fractional amplitudes (the upper one is burst 4 and the lower one is burst 7).  \label{fig:color_diagram}}
\end{figure}

Burst oscillations were also detected in burst 6. Oscillations in this case are detected in a 3 s interval during the tail of the burst and only in the 0.3--6.2 keV band. The fractional rms amplitude is $7.7 \pm 1.5 \%$, consistent with previous measurements of burst oscillation amplitudes from this source. Comparing the oscillations in the tail of these three bursts demonstrates that burst oscillations can be stronger in different energy bands and they can have different amplitudes and properties.

Prior work has shown that the properties of burst oscillations in 4U~1728 are correlated with the source spectral state, as evidenced by its position in the X-ray color-color diagram \citep{Franco2001,vanStraaten2001}. Indeed, burst oscillation amplitudes were shown to be largest at the high inferred accretion rates on the so-called ``banana branch." In an attempt to gain further insight into the high-amplitude oscillations we also explored their relationship to the source spectral state. Fig.~\ref{fig:Light_curve} shows the light curve and intensity vs.\ X-ray color diagrams for all of the \nicer{} data from 4U~1728 up to ObsID 1050150161 that ended at MJD 58313.14. The type I X-ray bursts are marked on these plots with red or orange symbols. The bursts with oscillations are shown in red. The two open circles show the bursts with high fractional oscillation amplitudes, and the filled circle shows the burst with low fractional amplitude. The soft color here is defined as the ratio of 1.8--2.5 keV photons over 0.5--1.8 keV photons, and the hard color is defined as the ratio of 3.8--6.8 keV over 2.5--3.8 keV. The count rate in this plot is computed in the 0.5--6.8 keV band. We also show the color-color diagram in Fig.~\ref{fig:color_diagram}. The definition of colors and the symbols are similar to Fig.~\ref{fig:Light_curve}.   Because the \nicer{} energy band differs from that of {\it RXTE}, we lack sufficient context to accurately determine which source states are present in Figure 6, and with only a modest number of burst oscillation detections it is difficult to draw any firm conclusions based on their locations in the figure. At present we do not see a clear indication that the high amplitude burst oscillations are source state dependent.

\acknowledgments

This work was supported by NASA through the \nicer{} mission and the 
Astrophysics Explorers Program. This research has made use of data and 
software provided by the High Energy Astrophysics Science Archive Research 
Center (HEASARC), which is a service of the Astrophysics Science Division at 
NASA/GSFC and the High Energy Astrophysics Division of the Smithsonian 
Astrophysical Observatory. DA thanks the Royal Society.

%

\vspace{5mm}
\facilities{\nicer{}}, ADS, HEASARC


\software{
HEAsoft (\url{https://heasarc.nasa.gov/lheasoft/})
}

\bibliographystyle{aasjournal}
\bibliography{nicer_burst}



\end{document}